\documentclass[%
 reprint,
 amsmath,amssymb,
 aps,
]{revtex4-2}

\usepackage{graphicx}
\usepackage{dcolumn}
\usepackage{bm}
\usepackage{xcolor}

\begin{document}

\preprint{APS/123-QED}

\title{Nonlocal discrete time crystals in periodically driven surface codes}

\author{Raditya Weda Bomantara}
\email{Raditya.Bomantara@sydney.edu.au}
\affiliation{%
 Centre for Engineered Quantum Systems, School of Physics,
University of Sydney, Sydney, New South Wales 2006, Australia
}%

\date{\today}

\begin{abstract}
Discrete time crystals (DTCs) are nonequilibrium phases of matter characterized by robust subharmonic order parameter dynamics. We report a new type of DTC in a periodically driven surface code, the subharmonic signature of which is only observable with respect to some nonlocal order parameter. By modifying the commonly used metrics for characterizing ordinary DTCs, i.e., the spectral functions, spin-glass order parameters, and two-point correlators, we further numerically demonstrate the rich phases of the proposed system. Specifically, depending on the system parameters and boundary conditions, we find that the system may fall into either a ``surface code" phase, trivial paramagnet phase, nonlocal period-doubling DTC phase, or nonlocal period-quadrupling DTC phase. Our work thus demonstrates the prospect of exploring topological codes for discovering novel phases of matter. 
\end{abstract}

\maketitle

\section{Introduction} 
Since the seminal work of Refs.~\cite{TC1,TC2}, the notion of the time crystals has intrigued many researchers and opened a new area in quantum many-body phases. The time crystals, if exist, demonstrate an instance of phases of matter with spontaneously broken time-translational symmetry and manifest themselves as a time-dependent ground state of an otherwise static and local Hamiltonian. While such continuous time crystals \cite{TC1,TC2} turn out to be impossible to exist \cite{NG1,NG2,NG3}, their nonequilibrium counterpart, termed discrete time crystals (DTCs), have been successfully achieved in experiments \cite{DTCexp1,DTCexp2,DTCexp3,DTCexp4,DTCexp5,DTCexp6,DTCexp7} following the theoretical proposals of Refs.~\cite{DTC1,DTC2,DTC3}. Such DTCs are defined on periodically driven many-body systems and feature a (local) order parameter, the dynamics of which is locked at a fraction of the driving frequency. In the last few years, various aspects of DTCs have been extensively studied \cite{DTC4,DTC5,DTC6,DTC7,DTC8,DTC9,DTC10,DTC11,DTC12,DTC13,DTC14,DTC15,DTC16,DTC17,DTC18,DTC19,DTC20,DTC21,DTC22,DTC23,DTC24,DTC25,DTC26,DTC27,DTC29,DTC30,DTC31,DTC32}, including their potential application in quantum computing \cite{DTCqc}, quantum simulation \cite{DTCqs,DTCqs2}, and the observation of condensed matter phenomena in the time domain \cite{DTCcm1,DTCcm2,DTCcm3,DTCcm5,DTCcm6}.

In a recent work by the same author \cite{DTCQEC}, the physics of DTCs is known to be closely related to a type of quantum error correction (QEC) code \cite{QEC1,QEC2,QEC7,QEC8}. Specifically, the observed robust period-doubling magnetization dynamics in periodically driven spin-1/2 DTCs originates from the exponential suppression of physical error with the system size due to the spin-spin interaction, which forms the stabilizer generators (mutually commuting operators) of the so-called quantum repetition code \cite{Rep1,Rep2,Rep3}. As the name suggests, the two logical basis states of the quantum repetition code are simply the tensor product of $|0\rangle$ and $|1\rangle$ states respectively, both of which represent the simultaneous eigenstates of all stabilizer generators with $+1$ eigenvalue. The presence of bit-flip errors changes the eigenvalue of some stabilizer generators from $+1$ to $-1$, which can in turn be detected and corrected. In the context of DTCs, these stabilizer generators appear at the hardware level as terms in the system's Hamiltonian. As such, DTCs are capable of performing passive error correction by { quasi-energetically} penalizing bit-flip error events. 

One main weakness of the quantum repetition code is its inability to correct phase-flip errors. While Ref.~\cite{DTCQEC} has shown that phase-flip errors do not break DTCs, the sensitivity of the quantum repetition code to phase-flip errors implies the presence of local order parameter exhibiting subharmonic dynamics. Indeed, a natural candidate for such order parameter is the logical $Z$ operator, which is capable of distinguishing the two logical states $|0\rangle$ and $|1\rangle$. In the repetition-code-based DTCs, this logical $Z$ operator is simply the average magnetization, which is also commonly used to probe period-doubling structure in previous literature.            

In this paper, we propose a new type of DTC based on the so-called surface code \cite{QEC4,QEC5,QEC6,QEC9,QEC10}, i.e., a more sophisticated topological QEC code defined on a two-dimensional square lattice. Unlike quantum repetition code, surface code is capable of correcting both bit-flip and phase-flip errors due to its nonlocal logical $X$ and $Z$ operators as strings of physical $X$ and $Z$ operators respectively, the weight of which grows with the system size. As an immediate consequence, the proposed DTC possesses a nonlocal order parameter displaying subharmonic dynamics and is hence termed nonlocal DTC onwards. To elucidate an explicit physical system supporting this exotic DTC, we consider a periodically driven surface code Hamiltonian and characterize its phases using a nonlocal generalization of some metrics previously used in studies of repetition-code-based DTCs \cite{DTC5}. Under open boundary conditions, we find that such a system may support either a ``surface code" spin-glass phase, which is topologically equivalent to a static surface code Hamiltonian, a nonlocal period-doubling DTC phase, or a trivial paramagnet phase. Interestingly, under periodic boundary conditions, the system may additionally support a nonlocal period-quadrupling DTC phase.   

This paper is organized as follows. In Sec.~\ref{sec:level1}, we introduce our system and elucidate the mechanism governing its nonlocal DTC phase. In Sec.~\ref{sec:level2}, we introduce a generalization of the three metrics previously used in studies of ordinary DTCs, i.e., the spectral functions, spin-glass order parameters, and two-point correlators, then utilize them to characterize the phases of the proposed system. In Sec.~\ref{sec:level3}, we discuss the effect of periodic boundary conditions on the system's phases, as well as some proposals for experimentally realizing the proposed system. We conclude this paper and present avenues for future studies in Sec.~\ref{sec:level5}.

\section{\label{sec:level1} Periodically driven surface code and $2T$-periodic nonlocal DTCs}
\begin{center}
\begin{figure}
    \includegraphics[scale=0.3]{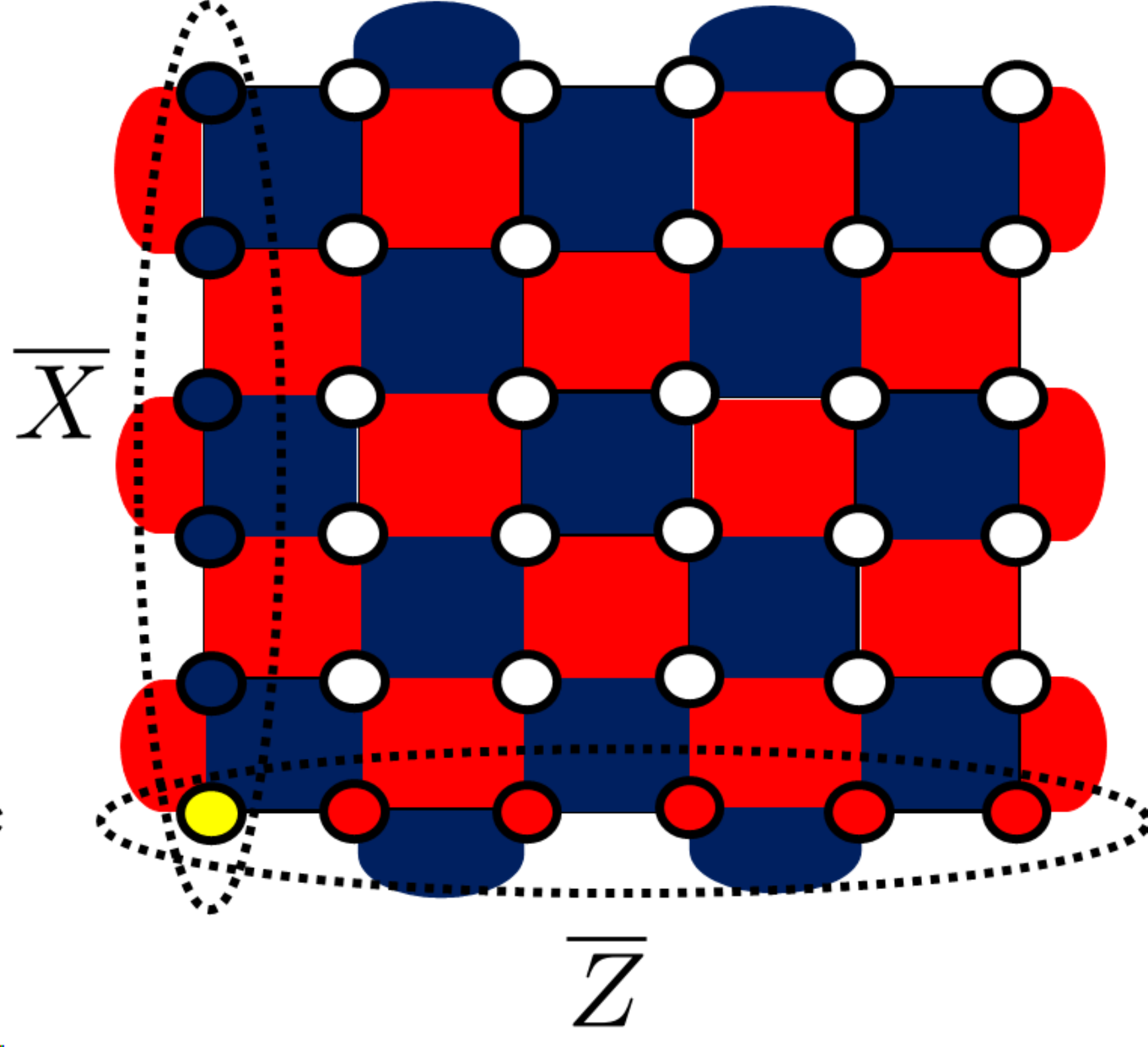}
    \caption{Schematics of $H_{S,\rm surface}$ on a $6\times 6$ lattice. Physical qubits live on the vertices of the lattice, $X$- and $Z$-type stabilizer generators are highlighted by the blue and red colors respectively, and the logical operators $\overline{X}$ and $\overline{Z}$ are highlighted by blue and red-filled circles respectively.}
    \label{fig:picsur}
\end{figure}
\end{center}

Consider a three-time-step Hamiltonian that gives rise to the one-period propagator (Floquet operator \cite{Flo1,Flo2}) of the form
\begin{equation}
    U_T = e^{\left(-\mathrm{i} \sum_{i,j} h_{x,i,j} X_{i,j} \right)} \times e^{\left(-\mathrm{i} \sum_{i,j} h_{z,i,j} Z_{i,j} \right)} \times  e^{\left(-\mathrm{i} H_{S,\rm surface}\right)} \;, \label{sys}
\end{equation}
{ where $h_{x,i,j}$ and $h_{z,i,j}$ can be understood as the magnetic field strength in the $x$- and $z$-direction respectively,} $H_{S,\rm surface}$ comprises the sum of the stabilizer generators associated with a $L_x\times L_y$ surface code, i.e.,
\begin{eqnarray}
H_{S,\rm surface} &=& \sum_{i=1}^\frac{L_x}{2} \sum_{j=1}^\frac{L_y}{2} \left(J_{2i-1,2j-1}\mathcal{S}_{X,2i-1,2j-1} +J_{2i,2j} \mathcal{S}_{X,2i,2j} \right. \nonumber \\
&& +\left. J_{2i-1,2j} \mathcal{S}_{Z,2i-1,2j} +J_{2i,2j-1} \mathcal{S}_{Z,2i,2j-1} \right) \nonumber \\
&& +\sum_{i=1,L_x}\sum_{j=1}^{\frac{L_y}{2}} J_{Z,i,2j-1} \mathcal{S}_{Z,i,2j-1}^{\rm boundary} \nonumber \\
&&+ \sum_{i=1}^{\frac{L_x}{2}} \sum_{j=1,L_y} J_{X,2i,j} \mathcal{S}_{X,2i,j}^{\rm boundary} \;, \label{surham}
\end{eqnarray}
{ $J_{i,j}, J_{Z,i,j}, J_{X,i,j}\in \mathbb{R}$ are the interaction strength,} $\mathcal{S}_{P,i,j}=P_{i,j}P_{i+1,j}P_{i,j+1}P_{i+1,j+1}$ for $P=X,Z$, $\mathcal{S}_{Z,i,j}^{\rm boundary}=Z_{i,j}Z_{i,j+1}$, and $\mathcal{S}_{X,i,j}^{\rm boundary}=X_{i,j}X_{i+1,j}$. Here, we assume that $L_x$ and $L_y$ are even; Similar Hamiltonian with appropriate modification of the boundary terms can be straightforwardly obtained if either $L_x$ or $L_y$ is odd. The various terms appearing in Eq.~(\ref{surham}) are schematically illustrated in Fig.~\ref{fig:picsur} with different colors. 

{ In the remainder of this paper, unless otherwise stated, all system parameters are drawn from a uniform distribution of the form $(\overline{P} -\Delta P, \overline{P} +\Delta P)$. Such a disorder is necessary for establishing many-body localization \cite{MBL1,MBL2,MBL3,MBL4} that prevents heating to a trivial infinite temperature state \cite{DTC2}. While alternative mechanisms to establishing long-lived DTCs without disorder exist in literature, such as prethermalization \cite{DTC3,pretherm2}, critical dynamics \cite{DTC7}, and all-to-all interaction \cite{DTC9}, their generalization to the nonlocal DTC setting is beyond the scope of this work.}

At the parameter values $h_{z,i,1}=h_{x,1,j}=\pi/2$ and $h_{z,i,j\neq 1}=h_{x,i\neq 1,j}=0$ (to be referred to as ideal values onwards), the first and second exponentials of Eq.~(\ref{sys}) reduce to perfect logical $Z$ and $X$ gates respectively, i.e., $\overline{Z}=\prod_{i=1}^{L_x} Z_{i,1}$ and $\overline{X}=\prod_{j=1}^{L_y} X_{1,j}$ in Fig.~\ref{fig:picsur}. In this case, by denoting $|\overline{0}\rangle$ and $|\overline{1}\rangle$ as the logical basis states in the simultaneous $+1$ eigenspace of all the stabilizer operators $\mathcal{S}_{X,i,j}$, $\mathcal{S}_{Z,i,j}$, $\mathcal{S}_{X,i,j}^{\rm boundary}$, and $\mathcal{S}_{Z,i,j}^{\rm boundary}$, one may explicitly construct two of the system's Floquet eigenstates as
\begin{equation} 
|\varepsilon_0, \pm\rangle = \frac{|\overline{0}\rangle \pm \mathrm{i} |\overline{1}\rangle}{\sqrt{2}} \;,
\end{equation}
with quasienergies 
\begin{eqnarray}
\varepsilon_{0,+}&=& \sum_{j,i} J_{i,j} +\sum_j \left(J_{Z,1,2j-1} +J_{Z,L_x,2j}\right) \nonumber \\
&& +\sum_i \left( J_{X,2i,1} +J_{X,2i,L_y} \right)
\end{eqnarray}
and $\varepsilon_{0,-}=\varepsilon_{0,+}+\pi/T$. In particular, the state $|\overline{0}\rangle$ transforms into $|\overline{1}\rangle$ over one period and vice versa, leading to a $2T$-periodicity capturable by the operator $\overline{Z}$. While $|\overline{0}\rangle$ and $|\overline{1}\rangle$ are highly entangled states, it can further be shown that $2T$-periodicity persists with respect to product states that can be written as superposition of logical states belonging to different stabilizer subspaces. An example of such a state is $|0\cdots 0 \rangle$, which can also be written as
\begin{eqnarray}
    |0\cdots 0\rangle &\propto & \sum_{s_{Z,k}, s_{X,k}=\pm 1} (1+\overline{Z}) \nonumber \\
    &\times&  \left(\prod_k (1+s_{Z,k}\mathcal{S}_{Z,k}) (1+s_{X,k}\mathcal{S}_{X,k}) \right) |0\cdots 0 \rangle \;, \nonumber \\
\end{eqnarray}
where $\left\lbrace \mathcal{S}_{Z,k}, \mathcal{S}_{X,k} \right\rbrace$ is the set of all $Z$ and $X$ stabilizers (including boundary operators).

Away from the ideal values above, the system is no longer integrable. Nevertheless, one may write the Floquet operator as
\begin{equation}
    U_T = \tilde{U}_T \times  U_{T,\rm ideal} \;, \label{sys2}
\end{equation}
where $U_{T,\rm ideal}$ is the Floquet operator associated with the ideal values,
\begin{equation}
    \tilde{U}_T = e^{\left(-\mathrm{i} \sum_{i,j} \delta_{x,i,j} X_{i,j} \right)} \times e^{\left(-\mathrm{i} \sum_{i,j} \delta_{z,i,j} Z_{i,j} \right)} \;,
\end{equation}
$\delta_{z,i,j} $ and $\delta_{x,i,j} $ are deviations in $h_{z,i,j}$ and $h_{x,i,j}$ from their respective ideal values. The evolution of a logical state, e.g., $|\overline{0}\rangle$, after one period can then be written as
\begin{eqnarray}
    U_T |\overline{0} \rangle &=& \tilde{U}_T | \overline{1} \rangle \nonumber \\
    &=& \langle \varepsilon_{0,-} |  \tilde{U}_T | \overline{1} \rangle  |\varepsilon_{0,-}\rangle  + \langle \varepsilon_{0,+} |  \tilde{U}_T | \overline{1} \rangle  |\varepsilon_{0,+} \rangle \nonumber \\
    && +\sum_{\varepsilon\neq \varepsilon_{0,+}, \varepsilon_{0,-}} \langle \varepsilon |  \tilde{U}_T | \overline{1} \rangle |\varepsilon \rangle \;, \label{phasediff}
\end{eqnarray}
where $|\varepsilon\rangle $ is a Floquet eigenstate of $U_{T,\rm ideal}$. By expanding the exponentials of $\tilde{U}_T$ as superpositions of products of $X_{1,j}$ and $Z_{i,1}$, it follows that nonzero contributions to $\langle \varepsilon_{0,\pm} |  \tilde{U}_T | \overline{1} \rangle$ come from identity, which leads to a relative $\pi$ phase between $|\varepsilon_{0,\pm}\rangle$, as well as $\prod_j^{L_y} X_{1,j}$ or $\prod_i^{L_x} Z_{i,1}$, which only appear at order $\sin^{L_y}(\delta_x)$ or $\sin^{L_x}(\delta_z)$ respectively. Indeed, lower weight Pauli operators bring $|\overline{1}\rangle$ outside the logical subspace, resulting in a state of zero overlap with $|\varepsilon_{0,\pm}\rangle$. Consequently, an imperfection of order $\sin(\delta_x), \sin(\delta_Z)\sim \epsilon$ will only result in the relative phase deviation from $\pi$ by an order of $\epsilon^{\mathrm{min}(L_x,L_y)}$.  

\begin{center}
\begin{figure}
    \includegraphics[scale=0.5]{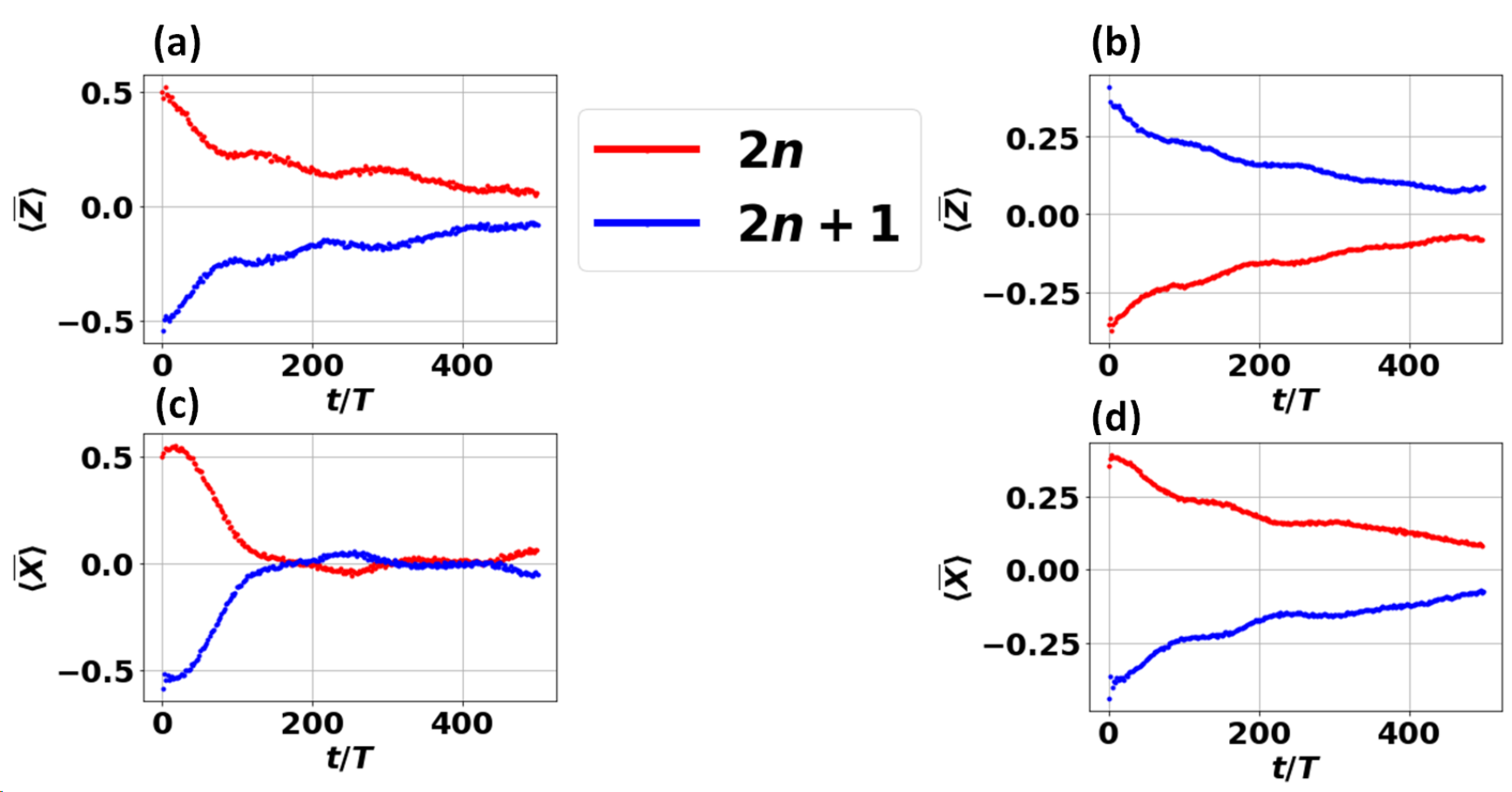}
    \caption{The stroboscopic dynamics of $\langle \overline{Z}\rangle$ and $\langle \overline{X}\rangle$ with respect to the initial state of $\prod_{i,j}e^{-\mathrm{i}\frac{\pi}{8}Y_{i,j}}|00\cdots 0\rangle$ in the surface code of size (a,c) $2\times 2$, (b,d) $3\times 3$. The system parameters are chosen as $\overline{h}_{x,1,j}=\overline{h}_{x,i\neq 1,j}-\pi/2=\overline{h}_{z,i,1}=\overline{h}_{z,i,j\neq 1}-\pi/2=0.45\pi$, $\overline{J}_{X,i,j}=\overline{J}_{Z,i,j}=0.3$, $\Delta h_{x,i,j}=\Delta h_{z,i,j}=0.005\pi$, and $\Delta J_{X,i,j}=\Delta J_{Z,i,j} = 0.5$, and each data point is averaged over $100$ disorder realizations.}
    \label{fig:pic1}
\end{figure}
\end{center}

Despite the similarity of the above feature with known DTC models based on quantum repetition code \cite{DTC2}, two main differences exist. First, a nonlocal order parameter, e.g., $\overline{Z}$, is necessary for capturing the $2T$ periodicity in the present system. Second, note that the usual DTC models typically only supports a single subharmonic order parameter dynamics, e.g., magnetization pointing in a specific direction. By contrast, at least two nonlocal observables with subharmonic dynamics exist in the present system, i.e., $\overline{Z}$ and $\overline{X}$. To support this point, Fig.~\ref{fig:pic1} displays the stroboscopic dynamics of $\overline{Z}$ and $\overline{X}$ for a surface code of two different sizes. There, robust $2T$ periodic profile is clearly evident in all cases despite the presence of considerable imperfection in the parameters $h_{x,j}$ and $h_{z,j}$ responsible for enacting the logical $X$ and $Z$ gate. 

\begin{center}
\begin{figure}
    \includegraphics[scale=0.5]{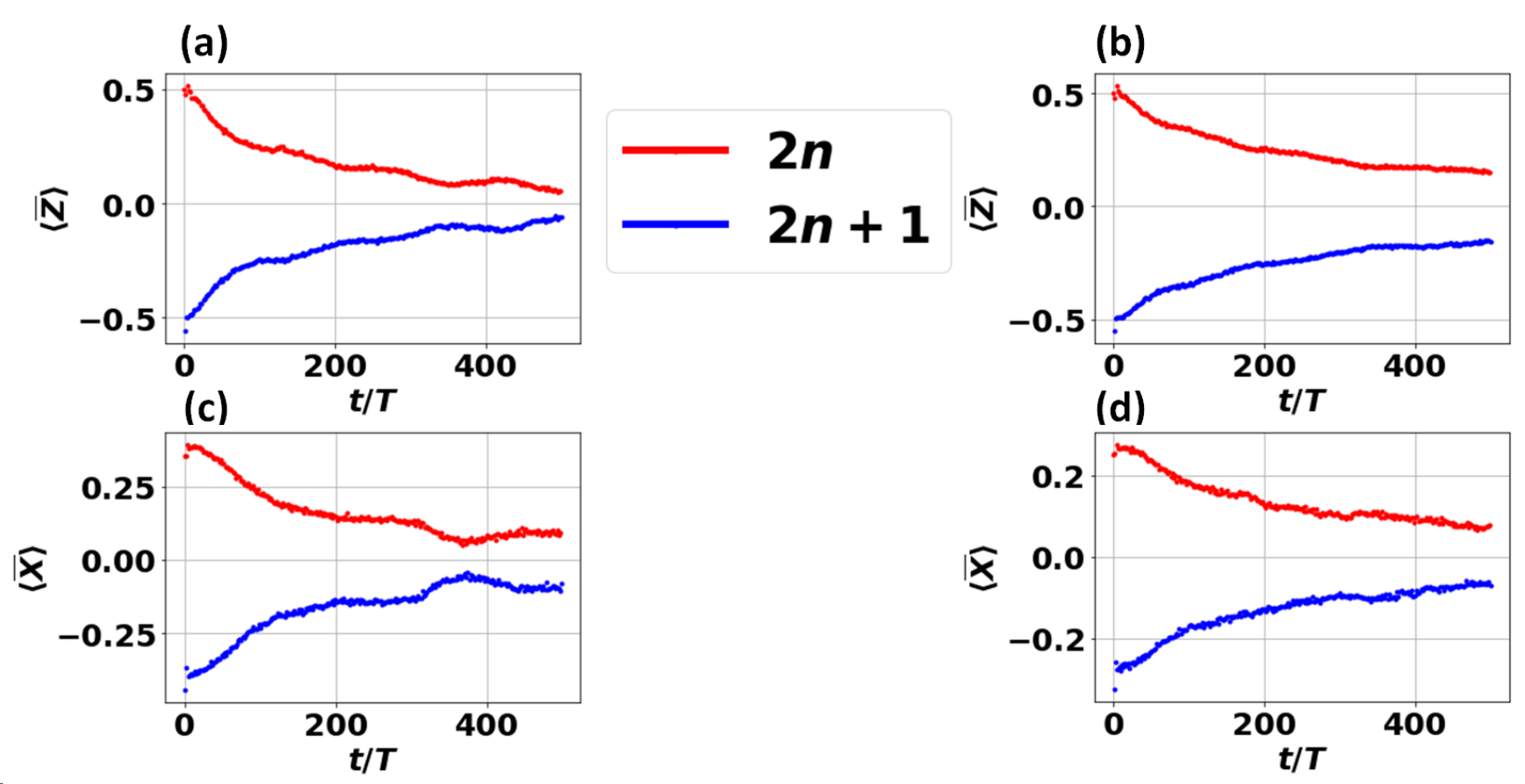}
    \caption{The stroboscopic dynamics of $\langle \overline{Z}\rangle$ and $\langle \overline{X}\rangle$ with respect to the initial state of $\prod_{i,j}e^{-\mathrm{i}\frac{\pi}{8}Y_{i,j}}|00\cdots 0\rangle$ in the surface code of size (a,c) $2\times 3$, (b,d) $2\times 4$. The system parameters are taken to be the same as those of Fig.~\ref{fig:pic1} and each data point is averaged over $100$ disorder realizations}
    \label{fig:pic4}
\end{figure}
\end{center}

Figure~\ref{fig:pic4} further reveals that increasing $L_y$ while fixing $L_x$ clearly improves the $2T$ oscillation profile of $\langle \overline{Z} \rangle$, but leaves that of $\langle \overline{X} \rangle$ to be qualitatively the same. This can be intuitively understood by inspecting the relative phase $\xi_{\overline{Z}} = \mathrm{i} \; \mathrm{log}\frac{\langle \varepsilon_{0,-}| \tilde{U}_T | \overline{1}\rangle}{\langle \varepsilon_{0,+}| \tilde{U}_T | \overline{1}\rangle}$ and $\xi_{\overline{X}} = \mathrm{i} \; \mathrm{log}\frac{\langle \varepsilon_{0,-}| \tilde{U}_T | \overline{+}\rangle}{\langle \varepsilon_{0,+}| \tilde{U}_T | \overline{+}\rangle}$ (recall Eqs.~(\ref{sys2})-(\ref{phasediff})), where $|\overline{+}\rangle =(|\overline{0} \rangle + |\overline{1} \rangle)/\sqrt{2}$. In particular, the initial state $|\overline{1} \rangle $ ($|\overline{+} \rangle$) leads to period-doubling in $\langle \overline{Z} \rangle$ ($\langle \overline{X} \rangle$) if $\xi_{\overline{Z}}\approx \pi$ ($\xi_{\overline{X}}\approx \pi$). By expanding $\tilde{U}_T$ as a superposition of products of $X_{i,j}$ and $Z_{i,j}$, it follows that the lowest order correction to $\xi_{\overline{X}}$ ($\xi_{\overline{Z}}$) from its ideal value of $\pi$ occurs at the $L_x$-($L_y$-)th order, which corresponds to terms $\propto \overline{Z}$ ($\propto \overline{X}$). This argument can be straightforwardly extended to any other eigenstate $|\psi_0\rangle $ of $H_{S,\rm surface}$. In this case, one may find two quasienergy eigenstates $|\varepsilon_{n,\pm}\rangle $ of $U_{T,\rm ideal}$ with $\pi/T$ spacing that satisfy $|\psi_0\rangle \propto |\varepsilon_{n,+}\rangle +|\varepsilon_{n,-}\rangle$. It then follows that $\mathrm{i} \; \mathrm{log}\frac{\langle \varepsilon_{n,-}| \tilde{U}_T | \psi_0\rangle}{\langle \varepsilon_{n,+}| \tilde{U}_T | \psi_0\rangle}\approx \pi$ up to a correction of order at least $\epsilon^{L_x}$ or $\epsilon^{L_y}$. Finally, by noting that any generic initial state can be expanded in terms of $U_{T, \rm ideal}$ eigenstates, one concludes that the deviation from a perfect period-doubling in $\langle \overline{X}\rangle $ ($\langle \overline{Z}\rangle $) is suppressed through an increase in $L_x$ ($L_y$), which agrees with the observation in Fig.~\ref{fig:pic4}.      

\section{Phase characterization} 
\label{sec:level2}


\subsection{Spectral functions}

We define a spectral function \cite{DTC5,pzmandppm,mypaper} associated with quasienergy $\epsilon$ and excitation operator $\psi$ as 
\begin{equation}
s_{\psi,\epsilon} = \mathcal{N}_{\psi,\epsilon} \sum_{n\in \mathcal{X}} \int_{-\delta}^{\delta} S_{\psi,\epsilon}(\varepsilon_n, \eta)   d\eta  \;, 
\end{equation} 
where $\epsilon\geq 0$, $\mathcal{X}$ is a set of some random integers smaller than the the system's Hilbert space dimension, $\delta\ll 1$, $\mathcal{N}_{\psi,\epsilon}=\frac{1}{\sum_{n\in \mathcal{X}}\int_{-\pi/T}^{\pi/T} S_{\psi,\epsilon}{\varepsilon_n,\eta} d\eta}$, and
\begin{eqnarray}
S_{\psi,\epsilon}(\varepsilon_n, \eta) &=& \sum_{k=-\infty}^{\infty} \sum_{\varepsilon_m} |\langle \varepsilon_n |\psi | \varepsilon_m \rangle |^2 \nonumber \\
&& \times \delta(\varepsilon_n-\varepsilon_m -\eta -\epsilon -2\pi k/T) \;.
\end{eqnarray}
Intuitively, $s_{\psi,\epsilon}$ measures the tendency of an operator $\psi$ to be a quasienergy $\epsilon$ excitation \cite{mypaper}. Indeed, if $s_{\psi,\epsilon}=1$, $\psi$ essentially maps any quasienergy eigenstate $|\varepsilon\rangle$ to $|\varepsilon+\epsilon\rangle$. 

In the system under consideration, the logical operators $\overline{Z}$ and $\overline{X}$ exactly represent zero or $\pi/T$ quasienergy excitations at two special parameter values. Specifically, in the ideal case, it can be easily verified that $\overline{Z}$ and $\overline{X}$ map any Floquet eigenstate $|\varepsilon_{n},\pm \rangle$ to another $|\varepsilon_{n},\mp \rangle$ that differs in quasienergy by $\pi/T$, i.e., $\overline{Z}$ and $\overline{X}$ are $\pi/T$ quasienergy excitations. On the other hand, at $h_{x,i,j}=h_{z,i,j}=0$, Eq.~(\ref{sys}) reduces to the time evolution of the surface code Hamiltonian and supports two-fold degenerate eigenstates. As logical operators, $\overline{Z}$ and $\overline{X}$ act within these degenerate eigenstate subspaces, mapping one Floquet eigenstate to another with the same quasienergy, i.e., $\overline{Z}$ and $\overline{X}$ thus represent zero quasienergy excitations. In general, it follows that the presence of $\pi/T$ quasienergy excitations yield $\pi/T$-separated quasienergy structure, and the system is consequently in a DTC phase. On the other hand, the presence of zero quasienergy excitations yield two-fold degenerate quasienergy structure, which is topologically equivalent to the surface code limit above. In this case, the system is termed to be in the ``surface code" phase. In the absence of zero and $\pi/T$ quasienergy excitations, the system is in the trivial phase.

To characterize the system's phase at general parameter values based on the presence or absence of zero and $\pi/T$ quasienergy excitations above, it is instructive to evaluate the four spectral functions $s_{\overline{Z},0}$, $s_{\overline{X},0}$, $s_{\overline{Z},\pi}$, and $s_{\overline{X},\pi}$. Our results are summarized in Fig.~\ref{fig:pic6}(a). There, the (nonlocal) DTC phase is associated with $s_{\overline{Z},\pi}=s_{\overline{X},\pi}\approx 1$ which, as expected, exists near the ideal values, i.e., $\overline{h}\approx \frac{\pi}{2}$. It is also worth pointing out the presence of ``surface code" phase, characterized by $s_{\overline{Z},0}=s_{\overline{X},0}\approx 1$ in the vicinity of the surface code limit $h_{x,i,j}=h_{z,i,j}=0$. Away from these special values, the system is in the trivial phase with $s_{\overline{Z},0}=s_{\overline{X},0}=s_{\overline{Z},\pi}=s_{\overline{X},\pi}\approx 0$. 


\subsection{Generalized spin-glass order parameters and nonlocal correlators}

\begin{center}
\begin{figure}
    \includegraphics[scale=0.5]{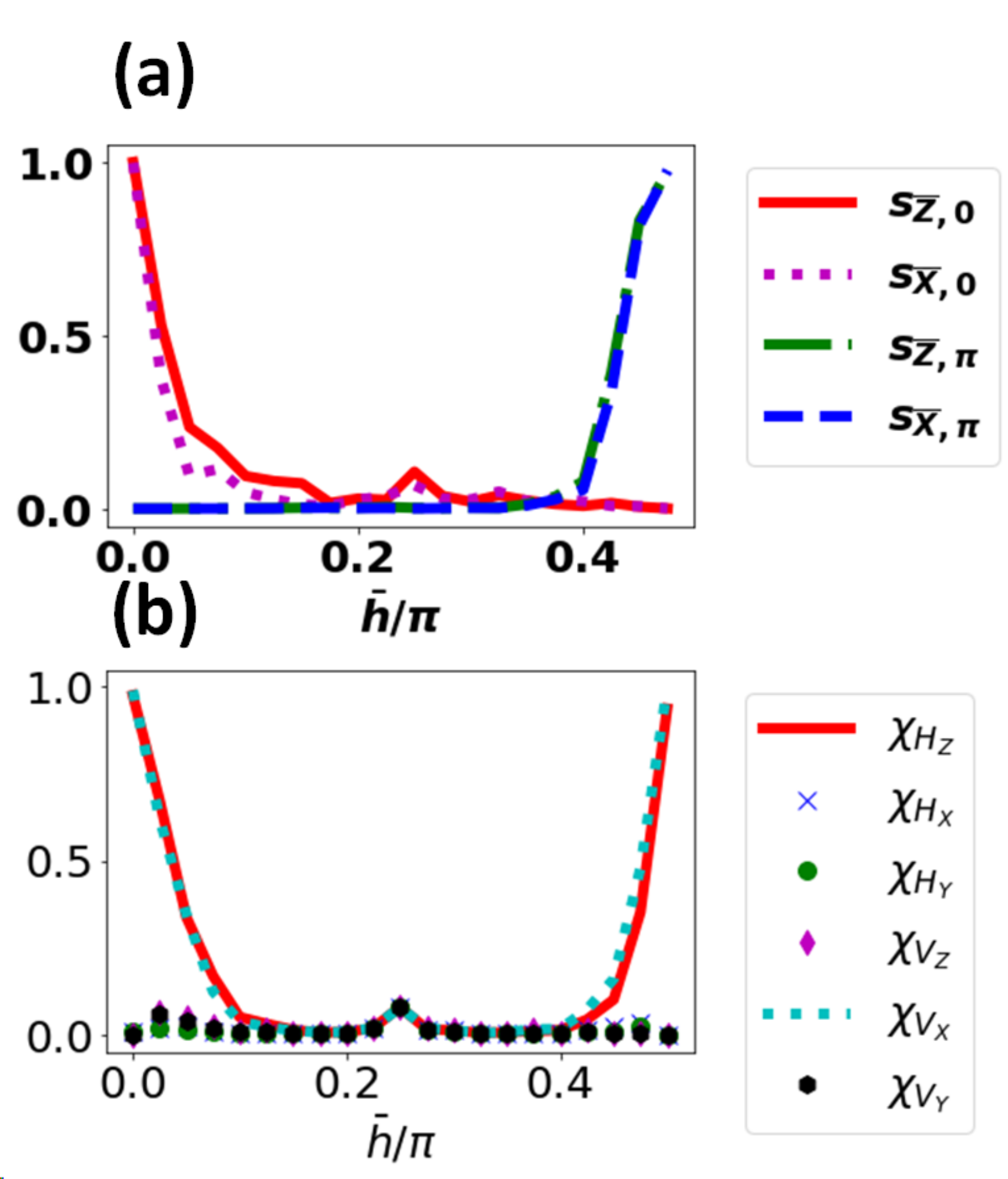}
    \caption{The (a) spectral functions and (b) generalized SG order parameters associated with the periodically driven surface code of size $3\times 3$. Apart from $\overline{h}_{x,1,j}=\overline{h}_{x,i\neq 1,j}-\pi/2=\overline{h}_{z,i,1}=\overline{h}_{z,i,j\neq 1}-\pi/2=\overline{h}$, other system parameters are the same as those of Fig.~\ref{fig:pic1} and each data point is averaged over $5$ disorder realizations.}
    \label{fig:pic6}
\end{figure}
\end{center}

Another useful diagnostic for characterizing the distinct phases in the system under consideration is to compute a set of generalized spin-glass (SG) order parameters defined as
\begin{eqnarray}
    \chi_{H_S}^{\rm (SG)} &=& \mathcal{N}_1 \sum_{i>j} \sum_{n=1}^D |\langle \varepsilon_n | H_{S,i} H_{S,j} | \varepsilon_n \rangle|^2 \;, \nonumber \\
    \chi_{V_S}^{\rm (SG)} &=& \mathcal{N}_2 \sum_{i>j} \sum_{n=1}^D |\langle \varepsilon_n | V_{S,i} V_{S,j} | \varepsilon_n \rangle|^2 \;, \nonumber \\
    \chi_{S}^{\rm (SG)} &=& \mathcal{N}_3 \sum_{(i,j) \neq (i',j')} \sum_{n=1}^D |\langle \varepsilon_n | S_{i,j} S_{i',j'} | \varepsilon_n \rangle|^2 \;, \nonumber \\
\end{eqnarray}
where $\mathcal{N}_1=\frac{2! (L_y^2-2)!}{(L_y^2)! D}$, $\mathcal{N}_2=\frac{2! (L_x^2-2)!}{(L_x^2)! D}$, $\mathcal{N}_3=\frac{ (L_y^2L_x^2-2)!}{(L_y^2L_x^2)! D}$, $H_{S,j}=\prod_{i=1}^{L_x} S_{i,j}$, $V_{S,j}=\prod_{j=1}^{L_y} S_{i,j}$, $S=X,Y,Z$, $D$ is the Hilbert space dimension, and $|\varepsilon_n\rangle$ is the $n$-th quasienergy eigenstate. It is worth pointing out that the order parameter $\chi_{S}^{\rm (SG)}$ has been previously employed in Ref.~\cite{DTC5} to distinguish between the SG and paramagnet (PM) phase in the one-dimensional periodically driven Ising chain. Interestingly, we found that $\chi_S^{\rm (SG)}\approx 0$ for any $S=X,Y,Z$ and parameter values in the periodically driven surface code Hamiltonian considered in this paper. This may naively suggest the absence of SG order in such a system, which clearly contradicts the robust bulk subharmonic dynamics observed in the previous section. On the other hand, as shown in Fig.~\ref{fig:pic6}(b), we found that $\chi_{H_Z}^{(SG)}$ and $\chi_{V_X}^{(SG)}$ are nonzero in the vicinity of $\overline{h}=0$ and $\overline{h}=\pi/2$, which correspond to the previously identified ``surface code" and nonlocal DTC phase respectively. This in turn implies the presence of a new type of nonlocal SG order.   

\begin{center}
\begin{figure}
    \includegraphics[scale=0.5]{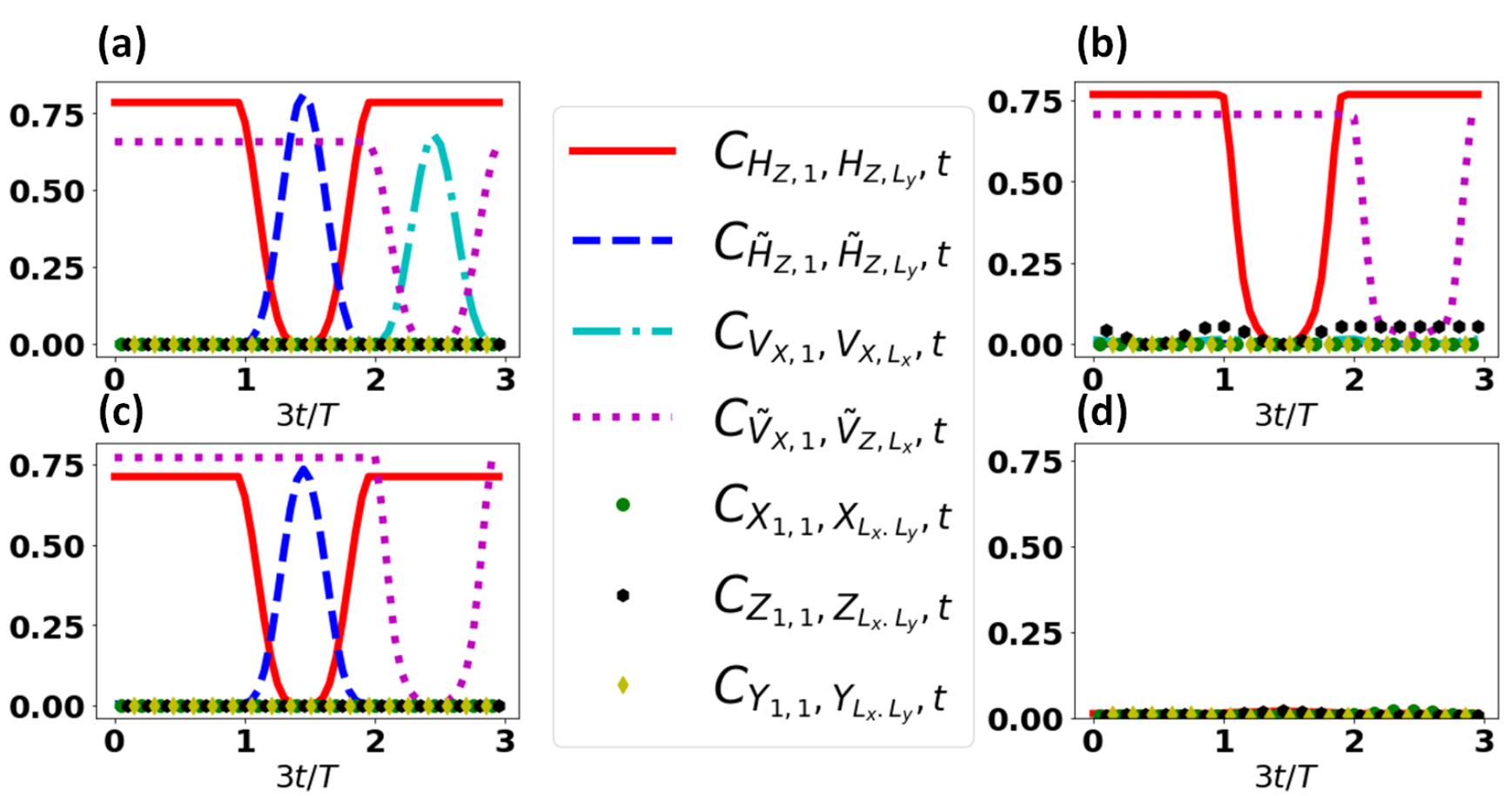}
    \caption{Nonlocal correlators' dynamics of $3\times 3$ periodically driven surface code Hamiltonian over a single period in the (a,c) nonlocal DTC regime, (b) surface code regime, and (d) trivial regime. The system parameters are taken as $\Delta J_{X,i,j}=\Delta J_{Z,i,j}=0.5$, $\Delta h_{X,i,j}=\Delta h_{Z,i,j}=0.0125\pi/2$  $\overline{J}_{X,i,j}=\overline{J}_{Y,i,j}=0.5$, (a) $\overline{h}_{X,1,j}=\overline{h}_{Z,i,1}=0.475\pi$, $\overline{h}_{X,i\neq 1,j}=\overline{h}_{Z,i,j\neq 1}=0.025\pi$, (b) $\overline{h}_{X,1,j}=\overline{h}_{Z,i,1}=0.025\pi$, $\overline{h}_{X,i\neq 1,j}=\overline{h}_{Z,i,j\neq 1}=0.475\pi$, (c) $\overline{h}_{X,1,j}=\overline{h}_{Z,i\neq 1,1}=0.475\pi$, $\overline{h}_{X,i\neq 1,j}=\overline{h}_{Z,1,j\neq 1}=0.025\pi$, and (d) $\overline{h}_{X,i,j}=\overline{h}_{Z,i,j}=0.225\pi$. All data points are averaged over 10 disorder realizations.}
    \label{fig:pic3}
\end{figure}
\end{center}

The nonlocal SG order in the ``surface code" and nonlocal DTC phase can be distinguished by further evaluating the correlators
\begin{eqnarray} 
    C_{H_{Z,i} H_{Z,j},t} &=& \langle \varepsilon_n (t) | H_{Z,i} H_{Z,j} |\varepsilon_n(t) \rangle \;, \nonumber \\
    C_{\tilde{H}_{Z,i} \tilde{H}_{Z,j},t} &=& \langle \varepsilon_n (t) | \tilde{H}_{Z,i} \tilde{H}_{Z,j} |\varepsilon_n(t) \rangle \;, \nonumber \\
    C_{V_{X,i} V_{X,j},t} &=& \langle \varepsilon_n (t) | V_{X,i} V_{X,j} |\varepsilon_n(t) \rangle \;, \nonumber \\
    C_{\tilde{V}_{X,i} \tilde{V}_{X,j},t} &=& \langle \varepsilon_n (t) | \tilde{V}_{X,i} \tilde{V}_{X,j} |\varepsilon_n(t) \rangle \;, \nonumber \\
    C_{S_{i,j},S_{i',j'},t} &=& \langle \varepsilon_n (t) | S_{i,j} S_{i',j'} |\varepsilon_n(t) \rangle \;, 
\end{eqnarray}
where $\tilde{H}_{Z,j}=Y_{1,j}\prod_{i=2}^{L_x} Z_{i,j}$, $\tilde{V}_{Z,i}=Y_{i,1}\prod_{j=2}^{L_y} Z_{i,j}$, $|\varepsilon_n(t)\rangle$ is a Floquet eigenstate at $t\in\left(0,T\right]$, $S=X,Y,Z$, and $|i-j|\gg 1$. In particular, as evidenced in Fig.~\ref{fig:pic3}, the DTC (surface code) phase is characterized by the presence (absence) of two crossing points between the nonlocal correlators $C_{H_{Z,1} H_{Z,L_y},t}$ and $C_{\tilde{H}_{Z,1} \tilde{H}_{Z,L_y},t}$ or $C_{V_{X,1} V_{X,L_x},t}$ and $C_{\tilde{V}_{X,1} \tilde{V}_{X,L_x},t}$ within $\left(0,T\right]$, whereas these nonlocal correlators are negligibly small in the PM phase. These crossing points signal the rotation of the SG order by $\pi$ with respect to either/both $\overline{X}$ or/and $\overline{Z}$, which in turn gives rise to the expected $2T$ observable dynamics in the nonlocal DTC regime. Finally, note that the local two-point correlators $C_{S_{i,j},S_{i',j'},t}\approx 0$ in all cases, further supporting our expectation that nonlocal operators are necessary to capture the system's nontrivial SG phases.   

\section{Discussion}
\label{sec:level3}




\subsection{Effect of boundary conditions}

A periodic boundaries variation of Eq.~(\ref{sys}) is obtained by modifying $H_{S,\rm surface}$ to
\begin{eqnarray}
H_{S,\rm toric} &=& \sum_{i=1}^\frac{L_x+2}{2} \sum_{j=1}^\frac{L_y+2}{2} \left(J_{2i-1,2j-1}\mathcal{S}_{X,2i-1,2j-1} +J_{2i,2j} \mathcal{S}_{X,2i,2j} \right. \nonumber \\
&& +\left. J_{2i-1,2j} \mathcal{S}_{Z,2i-1,2j} +J_{2i,2j-1} \mathcal{S}_{Z,2i,2j-1} \right) 
\end{eqnarray}
with $P_{i,j+L_y}=P_{i+L_x,j}=P_{i,j}$ for $P=X,Z$. That is, $H_{S,\rm toric}$ comprises a summation over the stabilizer generators of the paradigmatic toric code \cite{toric} defined on a torus geometry. 

\begin{center}
\begin{figure}
    \includegraphics[scale=0.5]{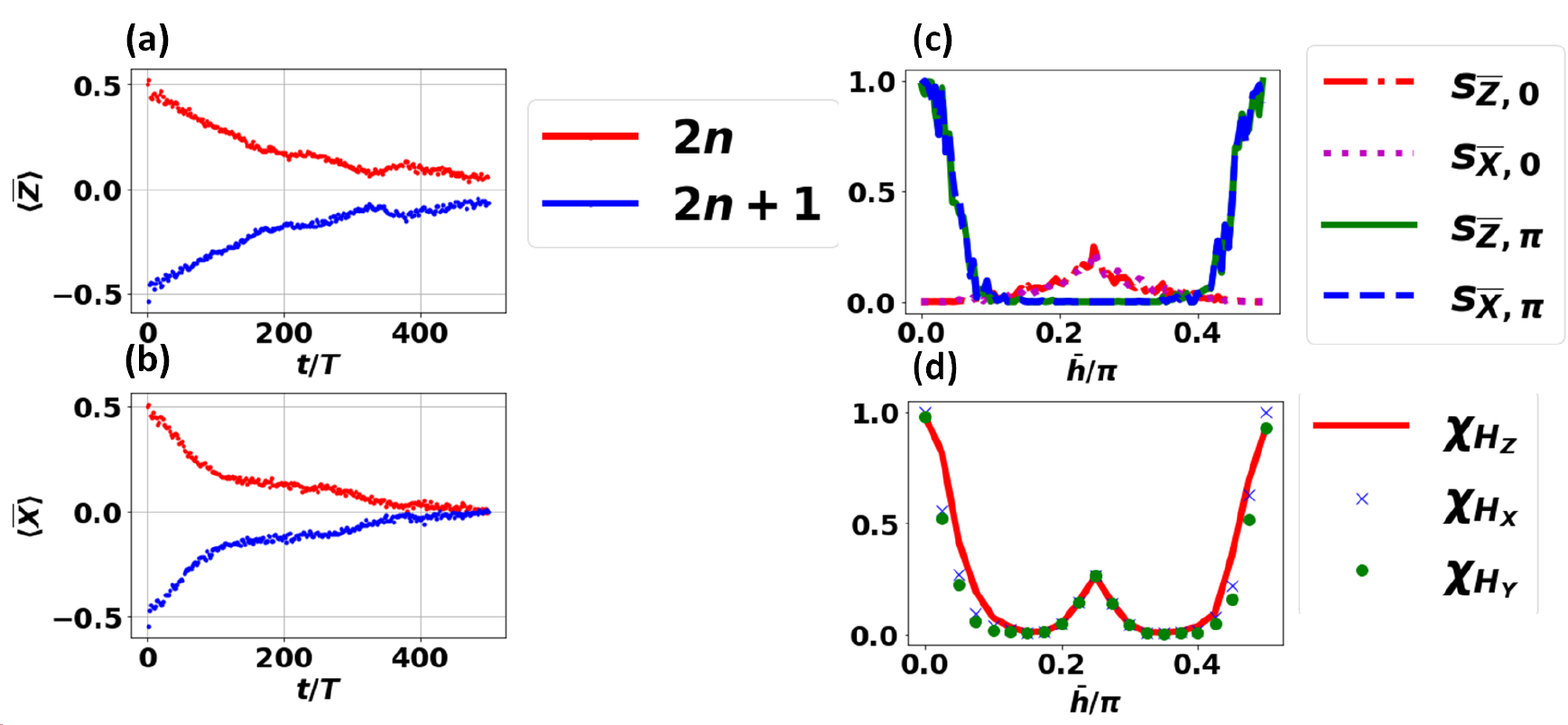}
    \caption{(a,b) The stroboscopic dynamics of $\overline{Z}\equiv \overline{Z}_1$ and $\overline{X} \equiv \overline{X}_1$ with respect to the initial state $\prod_{i,j}e^{\mathrm{i}\frac{\pi}{4}X_{i,j}}|00\cdots 0\rangle $ under the periodically driven toric Hamiltonian. The associated (c) spectral functions and (d) generalized SG order parameters as a function of the parameter $\overline{h}$. Other system parameters are taken to be the same as those in Fig.~\ref{fig:pic1} are used. The system size is taken to be (a,b,c) $2\times 2$ and (d) $2\times 4$. All data points are averaged over (a,b) $100$ and (c,d) $5$ disorder realizations.}
    \label{fig:pic8}
\end{figure}
\end{center}

Unlike the open boundaries surface code considered in the previous section, the toric code supports two logical qubits, i.e., $(\overline{X}_1,\overline{Z}_1)=(\prod_{j=1}^{L_y} X_{1,j},\prod_{i=1}^{L_x} Z_{i,1})$ and $(\overline{X}_2,\overline{Z}_2)=(\prod_{j=1}^{L_y} Z_{1,j},\prod_{i=1}^{L_x} X_{i,1})$. At the parameter values $h_{z,i,1}=h_{x,1,j}=\pi/2$ and $h_{z,i,j\neq 1}=h_{x,i\neq 1,j}=0$, each of the system's Floquet quasienergies is twofold degenerate. In particular, in terms of the logical basis states $|\overline{00}\rangle$ and $|\overline{11}\rangle$, a pair of degenerate quasienergy eigenstates can be explicitly constructed as
\begin{eqnarray}
    |\varepsilon_{0,1,\pm}\rangle &=& \frac{|\overline{00}\rangle \pm \mathrm{i} |\overline{10}\rangle }{\sqrt{2}} \;, \nonumber \\
    |\varepsilon_{0,2,\pm}\rangle &=& \frac{|\overline{01}\rangle \pm \mathrm{i} |\overline{11}\rangle }{\sqrt{2}} \;,
\end{eqnarray}
with quasienergies $\varepsilon_{0,1,+}=\varepsilon_{0,2,+}=\sum_{i,j} J_{i,j}$ and $\varepsilon_{0,1,-}=\varepsilon_{0,2,-}=\sum_{i,j} J_{i,j} +\pi/T$. It is to be emphasized however that the presence of such a degeneracy is not detrimental to the generation of subharmonic dynamics, since a perturbation lifting such a degeneracy (hence the $\pi/T$ spacing between pairs of quasienergies) must involve a nonlocal operator of either $\overline{X}_2$ or $\overline{Z}_2$. This is verified by explicitly computing the stroboscopic dynamics of $\langle \overline{Z}_1\rangle $ and $\langle \overline{X}_1\rangle $ in Fig.~\ref{fig:pic8}(a,b). For completeness, we also present the spectral functions and nonlocal order parameters in Fig.~\ref{fig:pic8}(c,d) as a function of the parameter $\overline{h}_{x,1,j}=\overline{h}_{x,i\neq 1,j}-\pi/2=\overline{h}_{z,i,1}=\overline{h}_{z,i,j\neq 1}-\pi/2=\overline{h}$. Of particular significance is the observation that the three nonlocal SG order parameters $\chi_{H_X}, \chi_{H_Y}, \chi_{H_Z}$ are nonzero in the SG regime. This feature of simultaneous SG order with respect to three nonlocal string operators is unique to the periodic boundaries setting and is made possible by the presence of additional logical operator $\overline{X}_2$.    

The presence of two logical qubits in the periodically driven toric code may further be exploited to find a parameter regime in which the system supports a $4T$-periodic DTC \cite{DTCQEC}. Specifically, at $h_{z,i,j}=\pi/4$, the second exponential in Eq.~(\ref{sys}) realizes a CNOT gate (up to an unimportant phase factor) with the first and second qubit being the control and target qubit respectively, i.e., it maps $\overline{X}_1\rightarrow \overline{X}_1 \overline{X}_2$, $\overline{Z}_1\rightarrow \overline{Z}_1$, $\overline{X}_2\rightarrow \overline{X}_2$, and $\overline{Z}_2\rightarrow \overline{Z}_1 \overline{Z}_2$. By further setting $h_{x,1,j}=\pi/2$ and $h_{x,i\neq 1, j}=0$, the system's Floquet operator reduces exactly to
\begin{equation}
    U_{T,\rm ideal}' = \overline{X}_1 \overline{CNOT}_{1,2} e^{-\mathrm{i} H_{S,toric}} \;,
\end{equation}
which maps $|\overline{00}\rangle \rightarrow |\overline{10}\rangle \rightarrow |\overline{01}\rangle \rightarrow |\overline{11}\rangle \rightarrow |\overline{00}\rangle$. One may then construct a quadruplet of $\pi/(2T)$ quasienergy separated eigenstates as
\begin{equation} 
    |\varepsilon_{0,\nu} \rangle = \frac{|\overline{00}\rangle + \nu |\overline{10}\rangle + \nu^2 |\overline{01}\rangle+ \nu^3 |\overline{11}\rangle}{2}\;, \label{4Teigen}
\end{equation}
where $\nu=1,\mathrm{i},-\mathrm{i},-1$. It follows that the system's other quasienergy eigenstates, which consist of states belonging to different stabilizer subspaces, also similarly form a quadruplet of $\pi/(2T)$ quasienergy separation. The $\pi/(2T)$ spacing in the system's quasienergy levels is in turn responsible for generating $4T$-periodic dynamics. Such $4T$-periodicity manifests itself, e.g., in the stroboscopic dynamics of $\overline{Z}_2$ with respect to the initial state $\prod_{i,j} e^{\mathrm{i}\frac{\pi}{4}X_{i,j}}|00\cdots 0\rangle $, which represents an equal weight superposition of $\overline{Z_1Z_2}=+1$ eigenstates over all stabilizer subspaces. 

\begin{center}
\begin{figure}
    \includegraphics[scale=0.5]{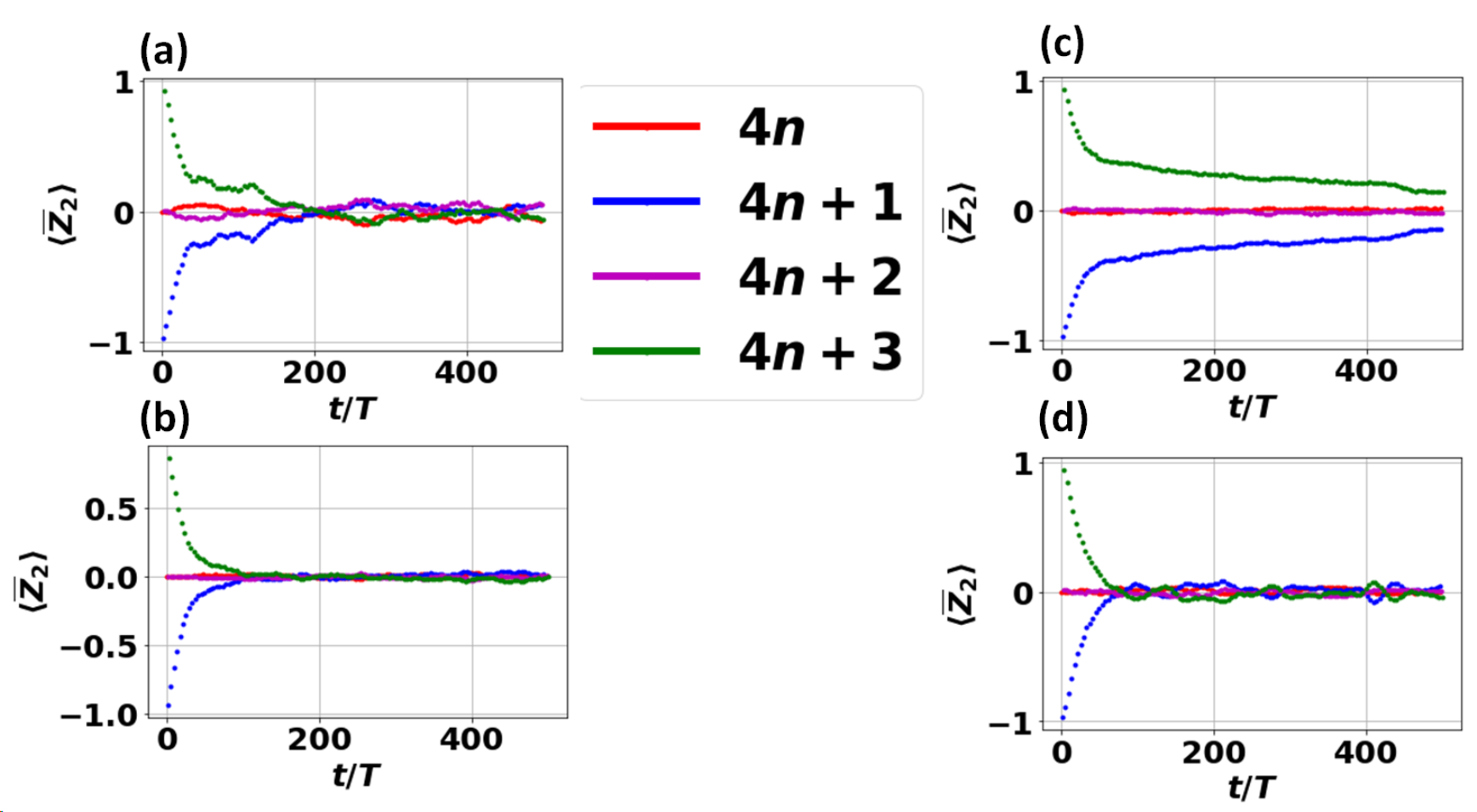}
    \caption{The stroboscopic dynamics of $\overline{Z}_2$ with respect to the initial state $\prod_{i,j}e^{\mathrm{i}\frac{\pi}{4}X_{i,j}}|00\cdots 0\rangle $ under the periodically driven toric Hamiltonian of size (a) $2\times 2$, (b) $4\times 2$, (c) $2\times 4$, and (d) under the periodically driven surface code (with open boundaries) of size $2\times 4$. We take $\Delta J_{X,i,j}=\Delta J_{Z,i,j}=0.5$, $\Delta h_{X,i,j}=0.0125\pi$, $\overline{J}_{X,i,j}=\overline{J}_{Y,i,j}=0.5$, and $\overline{h}_{X,1,j}=\overline{h}_{X,i\neq 1,j}-\pi/2=2\overline{h}_{Z,i,j}=0.475\pi$.}
    \label{fig:pic7}
\end{figure}
\end{center}

In Fig.~\ref{fig:pic7}, we plot $\langle \overline{Z_2}\rangle (t)$ slightly away from the special parameter values above to demonstrate the robust $4T$ oscillation expected from a DTC. Similar to the $2T$-periodic DTC setting, such robustness can be understood from the reduction of physical error of order $\epsilon$ to a logical error of order $\epsilon^{L_y}$ on some properly defined $\xi_{\overline{Z}_2}$ (See Appendix~\ref{app1} for detail). Intuitively, a logical error capable of destroying the system's period-quadrupling dynamics is either $\propto \overline{X}_2$ or $\propto \overline{X}_1$, which comprises a weight-$L_y$ string of $Z$ or $X$ operators respectively. This argument is confirmed in Fig.~\ref{fig:pic7}(c), which shows that an increase in $L_y$ significantly enhances and prolongs the observed $4T$ oscillation. By contrast, as demonstrated in Fig.~\ref{fig:pic7}(b), increasing $L_x$ does not seem to yield qualitative effect since the weight of $\overline{X}_1$ and $\overline{X}_2$ remains constant. Finally, as highlighted in Fig.~\ref{fig:pic7}(d), it is also worth noting that such $4T$-periodic DTC phase does not exist in the periodically driven surface code with open boundaries considered in the previous section.  

\subsection{Potential experimental realizations}

Ordinary DTCs have been realized in several experimental platforms, which include trapped ions \cite{DTCexp1,DTCexp7}, Nitrogen-Vacancy centers \cite{DTCexp2}, and nuclear magnetic moments \cite{DTCexp3,DTCexp4,DTCexp5}. Recently, the Sycamore qubit processor \cite{Sycamore} is identified to be another promising platform for simulating DTCs \cite{DTCqs2}. We expect that the periodically driven surface codes proposed in this paper can be approximated in trapped ions setup \cite{DTCexp1,DTCexp7}, or fully simulated with Sycamore device \cite{Sycamore} (or superconducting circuits in general).

In both trapped ions and Sycamore experiments, the weight-four interaction $\mathcal{S}_{X,i,j}$ and $\mathcal{S}_{X,i,j}$ of Eq.~(\ref{surham}) can be obtained by conjugating an appropriate weight-two qubit interaction with a combination of single- and two-qubit rotations. Specifically, the source of interaction in trapped ions experiments comes from the spin-dependent optical dipole forces that enact long-range Coulomb-like interaction of the form $G_{i,j}\equiv \sum_{i'\neq i} \frac{Z_{i,j}Z_{i',j}}{|i-i'|^\alpha}$ or $\tilde{G}_{i,j}\equiv \sum_{j'\neq j} \frac{Z_{i,j}Z_{i,j'}}{|j-j'|^\alpha}$ \cite{DTCexp1,DTCexp7}, where $0<\alpha<\infty$. Other types of weight-two interactions can then be obtained via appropriate single qubit rotations, e.g., $G_{i,j}^X \equiv \sum_j \frac{X_{i,j}Z_{i',j}}{|i-i'|^\alpha}=e^{-\mathrm{i} \frac{\pi}{4} Y_i } G_{i,j} e^{\mathrm{i} \frac{\pi}{4} Y_i}$, which can be realized through optically driven Raman transitions between two hyperfine clock states of $^{171}Yb+$ ion. In the limit $\alpha \rightarrow \infty$, $G_{i,j}$ and $\tilde{G}_{i,j}$ reduce exactly to the nearest-neighbor Ising interaction. One can then exactly produce $e^{-\mathrm{i} \mathcal{S}_{X,i,j}}$ and $e^{-\mathrm{i} \mathcal{S}_{Z,i,j}}$ by applying a series of conjugation on $e^{-\mathrm{i} G_{i,j}}$ (See Appendix~\ref{app2} for detail). At finite $\alpha$, the same prescription may still yield a $\mathcal{S}_{X,i,j}$ or $\mathcal{S}_{Z,i,j}$ term, albeit with a smaller amplitude, as well as unwanted higher-weight spin-spin interactions. In the experiment of Ref.~\cite{DTCexp1}, a maximum of $\alpha=3$ can in principle be achieved. In this case, one may either exploit the presence of these unwanted interactions as perturbation, or design a more sophisticated experimental scheme to remove them altogether, e.g., via a series of dynamical decoupling sequences.

\begin{center}
\begin{figure}
    \includegraphics[scale=0.5]{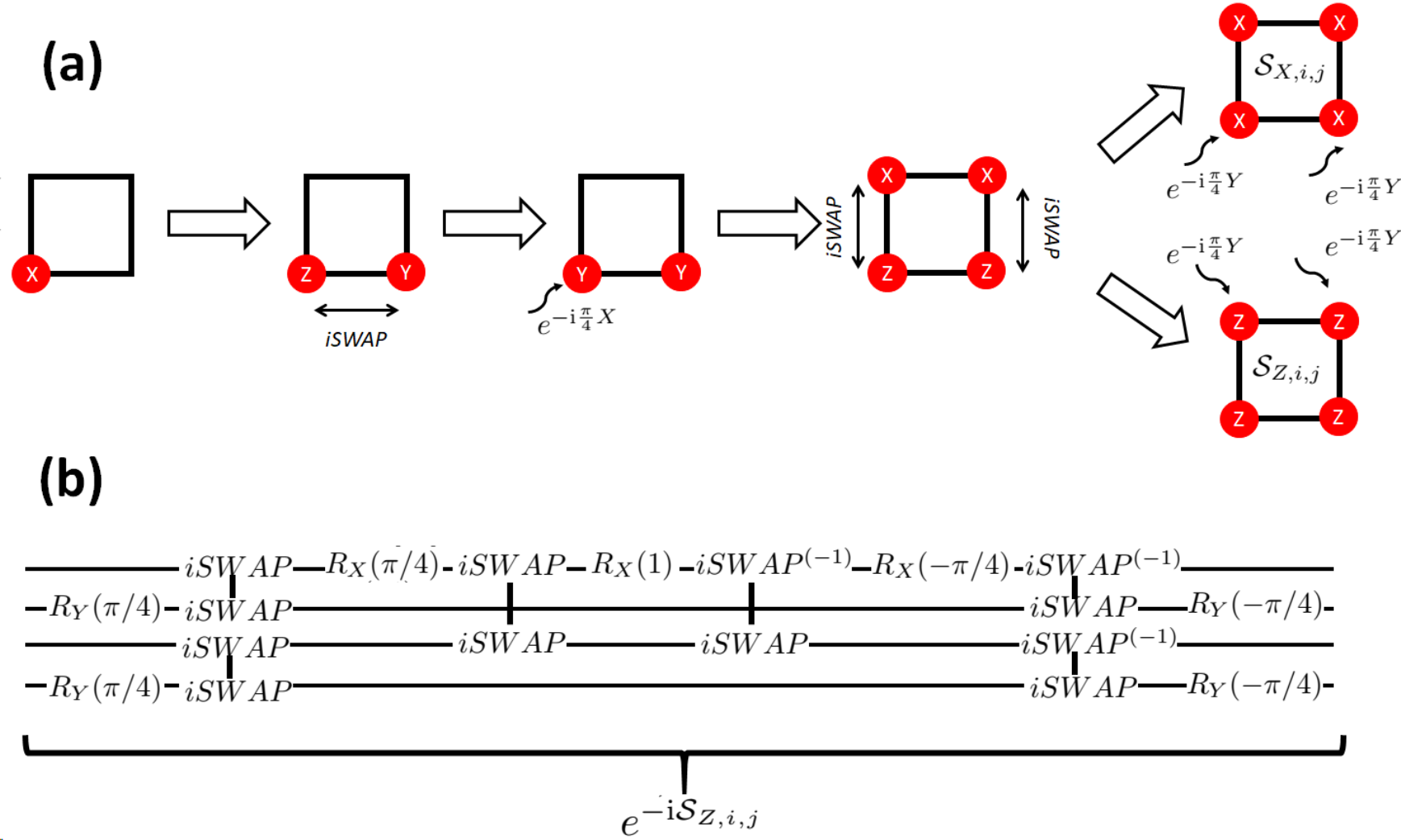}
    \caption{(a) Generation of $\mathcal{S}_{X,i,j}$ and $\mathcal{S}_{Z,i,j}$ by conjugating a single Pauli operator with a series of iSWAP and single qubit gates. (b) The associated quantum circuit implementation of $\mathcal{S}_{Z,i,j}$.}
    \label{fig:picsyc}
\end{figure}
\end{center}

The Sycamore device \cite{Sycamore} may prove to be a more suitable platform for exactly implementing the proposed model. To this end, the stabilizers $\mathcal{S}_{X,i,j}$ and $\mathcal{S}_{Z,i,j}$ terms can be implemented solely from a combination of single qubit rotations $R_P(\theta) \equiv \exp\left(-\mathrm{i} \theta P\right)$ ($P=X,Y,Z$) and $iSWAP_{(i,j),(i',j')}=e^{-\mathrm{i} \frac{\pi}{4} (X_{i,j}X_{i',j'}+Y_{i,j}Y_{i',j'})}$ gate (see Fig.~\ref{fig:picsyc} and Appendix~\ref{app2} for detail), both of which are native within the Sycamore platform \cite{Sycamore} and can be executed with a very high fidelity.

\section{Conclusion}
\label{sec:level5}

We have introduced a periodically driven surface code and highlighted its nonlocal DTC phase. Unlike other previously studied DTCs, a nonlocal order parameter is neccesary for capturing its subharmonic signature. The nonlocal generalization of the spectral functions, spin-glass order parameters, and two-point correlators are further introduced to distinguish such nonlocal DTCs from trivial paramagnets. These metrics further allow the characterization of the system's other phases, which include the trivial paramagnetic and ``surface code" phase. The latter is topologically equivalent to the static surface code system in the absence of magnetic field.

The nonlocal DTCs observed in this paper are also expected to exist in a variety of other topologically ordered systems. To this end, a comprehensive study of other periodically driven topological quantum error correction codes, e.g., the color codes \cite{color1,color2}, the Majorana version of the surface codes \cite{Majsur}, or the quantum double models \cite{toric}, thus represents a promising direction for future work. In this case, the three nonlocal metrics introduced above will further open an opportunity for discovering other exotic phases in these systems. To improve their experimental feasibility beyond trapped ions and Sycamore platforms, it will also be interesting to explore the possibility of observing nonlocal DTCs with continuous driving. Finally, given that nonlocal DTCs share similar physics as that of topological quantum error correction codes, their potential application in quantum computing poses another interesting open question worth exploring.       

{\it Note added:} During the finalization of this work, we came across a recent similar preprint \cite{similar} proposing a new type of discrete time crystal in periodically driven topological ordered systems, with a particular example on the periodically driven surface codes. Despite the similarity in the model under study, Ref.~\cite{similar} focuses on rigorously establishing the theory of topologically-ordered time crystals (equivalent to nonlocal DTCs introduced here), whereas this work focuses on characterizing the various phases of the system under both periodic and open boundary conditions.   

\begin{acknowledgements}
	This work is supported by the Australian Research Council Centre of Excellence for Engineered Quantum Systems (EQUS, CE170100009). The author thanks Isaac Kim for useful discussion that motivates the conception of this work.
\end{acknowledgements}

\onecolumngrid 
\appendix

\section{Robustness of period-quadrupling $\langle \overline{Z}_2 \rangle $ in the periodically driven toric code}
\label{app1} 

By writing $\delta_{z,i,j}=h_{z,i,j}-\pi/4$, $\delta_{z,1,j}=h_{x,1,j}-\pi/2$, and $\delta_{x,i\neq 1,j}=h_{x,i\neq 1,j}-\pi/2$, one obtains
\begin{eqnarray}
    U_T &=& e^{-\mathrm{i} \delta_{x,i,j} X_{i,j}} \overline{X}_1 e^{-\mathrm{i} \delta_{z,i,j} Z_{i,j}} \overline{CNOT}_{1,2} e^{-\mathrm{i} H_{S,\rm toric}} \nonumber \\
    &=& e^{-\mathrm{i} \delta_{x,i,j} X_{i,j}} e^{-\mathrm{i} \tilde{\delta}_{z,i,j} Z_{i,j}}  \overline{X}_1 \overline{CNOT}_{1,2} e^{-\mathrm{i} H_{S,\rm toric}} \equiv \tilde{U}_T U_{T,\rm ideal} \;, 
\end{eqnarray}
where $\tilde{\delta}_{z,1,j}=-\delta_{z,1,j}$ and $\tilde{\delta}_{z,i\neq 1,j}=\delta_{z,i\neq 1,j}$. Consider the initial state $|\psi\rangle =e^{\mathrm{i}\frac{\pi}{4}X_{i,j}}|00\cdots 0\rangle $. It satisfies $\overline{Z_1Z_2} |\psi\rangle = + |\psi\rangle$ and has uniform support on all $\overline{Z_1Z_2}=+1$ eigenstates belonging to different stabilizer subspaces. The latter can be immediately shown by writing
\begin{equation}
    |\psi\rangle \propto \sum_{s_{X,i,j},s_{Z,i,j}=\pm 1} (1+s_{X,i,j}\mathcal{S}_{X,i,j})(1+s_{Z,i,j}\mathcal{S}_{Z,i,j}) (1+\overline{Z_1Z_2}) |\psi\rangle\;, 
\end{equation}
up to some normalization constant, where it is recognized that
\begin{equation}
|\psi\rangle_{s_{X,i,j},s_{Z,i,j}} = (1+s_{X,i,j}\mathcal{S}_{X,i,j})(1+s_{Z,i,j}\mathcal{S}_{Z,i,j}) (1+\overline{Z_1Z_2}) |\psi\rangle =(1+s_{X,i,j}\mathcal{S}_{X,i,j})(1+s_{Z,i,j}\mathcal{S}_{Z,i,j}) |\psi\rangle   
\end{equation}
represents a $\overline{Z_1Z_2}=+1 $ eigenstate belonging to the $\mathcal{S}_{X,i,j}=s_{X,i,j}$ and $\mathcal{S}_{Z,i,j}=s_{Z,i,j}$ stabilizer subspace. Associated with each $|\psi\rangle_{s_{X,i,j},s_{Z,i,j}}$, one may find a quadruplet of $U_{T,\rm ideal}$ eigenstates $|\varepsilon_{s_{X,i,j},s_{Z,i,j},\nu}\rangle$ with $\pi/(2T)$ spacing such that
\begin{equation}
|\psi\rangle_{s_{X,i,j},s_{Z,i,j}}=\frac{|\varepsilon_{s_{X,i,j},s_{Z,i,j},1}\rangle + |\varepsilon_{s_{X,i,j},s_{Z,i,j},\mathrm{i}}\rangle +|\varepsilon_{s_{X,i,j},s_{Z,i,j},-\mathrm{i}}\rangle +|\varepsilon_{s_{X,i,j},s_{Z,i,j},-1}\rangle}{2} \;.    
\end{equation}
In the special case of $s_{X,i,j}=s_{Z,i,j}=+1$, these four eigenstates are precisely given by Eq.~(\ref{4Teigen}) in the main text. In other stabilizer subspaces, such eigenstates can be constructed via the superposition
\begin{equation}
 |\varepsilon_{s_{X,i,j},s_{Z,i,j},\nu}\rangle = \frac{1 + \nu U_{T,\rm ideal} +\nu^2 U_{T,\rm ideal}^2  +\nu^3 U_{T,\rm ideal}^3}{2} |\psi\rangle_{s_{X,i,j},s_{Z,i,j}} \;, \label{4Teigengen}     
\end{equation}
which can be directly verified by applying $U_{T,\rm ideal}$ directly on $|\varepsilon_{s_{X,i,j},s_{Z,i,j},\nu}\rangle$ and noting that $U_{T,\rm ideal}^4 |\psi\rangle_{s_{X,i,j},s_{Z,i,j}} \propto |\psi\rangle_{s_{X,i,j},s_{Z,i,j}}$. 

Taking into account parameter imperfection, the one-period evolution of $|\psi\rangle$ is immediately obtained as 
\begin{eqnarray}
    U_T |\psi\rangle &=& \sum_{s_{X,i,j},s_{Z,i,j},\nu} \langle \varepsilon_{s_{X,i,j},s_{Z,i,j},\nu } | \tilde{U}_T |\psi\rangle_{s_{X,i,j},s_{Z,i,j}}^{(1)} |\varepsilon_{s_{X,i,j},s_{Z,i,j},\nu }\rangle + \cdots \;,
\end{eqnarray}
where $|\psi\rangle_{s_{X,i,j},s_{Z,i,j}}^{(1)}=U_{T,\rm ideal} |\psi\rangle_{s_{X,i,j},s_{Z,i,j}} $ and $(\cdots)$ refers to all cross terms of the form $\langle \varepsilon_{s_{X,i,j},s_{Z,i,j},\nu } | \tilde{U}_T |\psi\rangle_{s'_{X,i,j},s'_{Z,i,j}}^{(1)} |\varepsilon_{s_{X,i,j},s_{Z,i,j},\nu }\rangle$ with $s'_{X,i,j}\neq s_{X,i,j}$ or $s'_{Z,i,j}\neq s_{Z,i,j}$. To demonstrate the robustness of the system's period-quadrupling feature, it suffices to show that $\langle \varepsilon_{s_{X,i,j},s_{Z,i,j},\nu } | \tilde{U}_T |\psi\rangle_{s_{X,i,j},s_{Z,i,j}}^{(1)}$ maintains a relative phase close to $\pi/(2T)$. The latter is immediately established by expanding $\tilde{U}_T$ as a superposition of products of $Z_{i,j}$ and $X_{i,j}$, then noting that the next nonzero contribution to $\langle \varepsilon_{s_{X,i,j},s_{Z,i,j},\nu } | \tilde{U}_T |\psi\rangle_{s_{X,i,j},s_{Z,i,j}}^{(1)}$ after identity, which establishes a $\pi/(2T)$ relative phase, occurs at order $\epsilon^{L_y}$ ($\epsilon\sim \mathrm{max}\left(|\sin(\delta_{x,i,j})|,|\sin(\delta_{z,i,j})|\right) $) and is either $\propto \overline{X}_1$ or $\propto \overline{X}_2$.  

\section{Low-weight qubit implementation of the stabilizer operators in trapped ions and superconducting circuits}
\label{app2}

Our construction relies heavily on the Euler-like identity
\begin{equation}
    e^{\mathrm{i} \theta \mathcal{P}_1} \mathcal{P}_2 e^{-\mathrm{i} \theta \mathcal{P}_1} = \cos(2\theta) \mathcal{P}_2 -\mathrm{i} \sin(2\theta) \mathcal{P}_1 \mathcal{P}_2 \;, \label{Eq:App1}
\end{equation}
where $\mathcal{P}_1$ and $\mathcal{P}_2$ are two anticommuting Pauli matrices. 

\subsection{Trapped ions}

For simplicity, we work in the limit $\alpha=\infty$ and denote $G_{i,j}=Z_{i,j} Z_{i+1,j}+Z_{i,j} Z_{i-1,j}$, $\tilde{G}_{i,j}=Z_{i,j} Z_{i,j+1}+Z_{i,j} Z_{i,j-1}$, $G_{i,j}^P=P_{i,j} Z_{i+1,j}+P_{i,j} Z_{i-1,j}$, and $\tilde{G}_{i,j}^P=P_{i,j} Z_{i,j+1}+P_{i,j} Z_{i,j-1}$ for $P=X,Y,Z$. To produce $e^{\mathrm{i} \mathcal{S}_{X,i,j}}$, one may start with $e^{-\mathrm{i} G_{i,j}}$, then conjugating it with $e^{\sum_{i'} \mathrm{i}  \frac{\pi}{4} X_{i',j} } $, followed by $e^{\sum_{i'} \mathrm{i}  \frac{\pi}{4} G_{i',j+1}^X } $. Indeed,

    \begin{eqnarray}
    e^{\sum_{i'} \mathrm{i} \frac{\pi}{4} X_{i',j} } e^{-\mathrm{i} G_{i,j}} e^{-\sum_{i'}\mathrm{i} \frac{\pi}{4} X_{i',j} } &=& e^{-\mathrm{i} (Y_{i,j}Y_{i+1,j}+Y_{i,j}Y_{i-1,j}) } \;, \nonumber \\
    e^{\sum_{i'} \mathrm{i}  \frac{\pi}{4} \tilde{G}_{i',j+1}^X } e^{-\mathrm{i} (Y_{i,j}Y_{i+1,j}+Y_{i,j}Y_{i-1,j}) } e^{-\sum_{i'} \mathrm{i}  \frac{\pi}{4} \tilde{G}_{i',j+1}^X } &=& e^{-\mathrm{i} (X_{i,j}X_{i,j+1}X_{i+1,j}X_{i+1,j+1}+X_{i,j}X_{i,j+1}X_{i-1,j}X_{i-1,j+1}) }  \;.
\end{eqnarray}
Stabilizer operators of the form $e^{\mathrm{i} \mathcal{S}_{Z,i,j}}$ can be produced by additionally applying appropriate $\pi/4$ Pauli-$Y$ rotation to the above scheme.   

\subsection{Superconducting circuits}

In a similar fashion, the implementation of the stabilizer operators with Sycamore's native gates prescribed in Fig.~\ref{fig:picsyc} of the main text can be directly verified by repeated applications of Eq.~(\ref{Eq:App1}). That is, 
\begin{eqnarray}
    iSWAP_{(i,j),(i+1,j)}^{(-1)} e^{-\mathrm{i} X_{i,j}} iSWAP_{(i,j),(i+1,j)} &=&  e^{-\mathrm{i} Z_{i,j} Y_{i+1,j}} \;, \nonumber \\
    e^{\mathrm{i} \frac{\pi}{4} X_{i,j}} e^{-\mathrm{i} Z_{i,j} Y_{i+1,j}} e^{-\mathrm{i} \frac{\pi}{4} X_{i,j}} &=& e^{-\mathrm{i} Y_{i,j} Y_{i+1,j}} \;, \nonumber \\
    iSWAP_{(i+1,j),(i+1,j+1)}^{(-1)} iSWAP_{(i,j),(i,j+1)}^{(-1)} e^{-\mathrm{i} Y_{i,j} Y_{i+1,j}} iSWAP_{(i,j),(i,j+1)} iSWAP_{(i+1,j),(i+1,j+1)} &=& e^{-\mathrm{i} Z_{i,j} X_{i,j+1} Z_{i+1,j} X_{i+1,j+1} } \;, \nonumber \\ 
    e^{\mathrm{i} \frac{\pi}{4} Y_{i+1,j+1}} e^{\mathrm{i} \frac{\pi}{4} Y_{i,j+1}} e^{-\mathrm{i} Z_{i,j} X_{i,j+1} Z_{i+1,j} X_{i+1,j+1} } e^{-\mathrm{i} \frac{\pi}{4} Y_{i,j+1}} e^{-\mathrm{i} \frac{\pi}{4} Y_{i+1,j+1}} &=& e^{-\mathrm{i} Z_{i,j} Z_{i,j+1} Z_{i+1,j} Z_{i+1,j+1} } \nonumber \\
    & \equiv & e^{-\mathrm{i} \mathcal{S}_{Z,i,j}} \;,
\end{eqnarray}
and similarly for $e^{-\mathrm{i} \mathcal{S}_{X,i,j}}$.

\end{document}